# Computer Models Design for Teaching and Learning using Easy Java Simulation

Loo Kang Lawrence WEE[1], Ai Phing LIM[2], Khoon Song Aloysius GOH[5], Sze Yee LYE[1], Tat Leong LEE[2], Weiming XU[2], Giam Hwee Jimmy Goh[3], Chee Wah ONG[4], Soo Kok NG[4], Ee-Peow LIM[5], Chew Ling LIM[6], Wee Leng Joshua YEO[6], Matthew ONG[1], Kenneth Y T LIM[7]

[1] Ministry of Education, Education Technology Division (ETD), Singapore
[2] Ministry of Education, River Valley High School (RVHS), Singapore
[3] Ministry of Education, Yishun Junior College (YJC), Singapore
[4] Ministry of Education, Innova Junior College (IJC), Singapore
[5] Ministry of Education, Anderson Junior College (AJC), Singapore
[6] Ministry of Education, Serangoon Junior College (SRJC), Singapore
[7] National Institute of Education, Nanyang Technological University, Singapore

wee_loo_kang@moe.gov.sg, lim_ai_phing@moe.edu.sg, goh_khoon_song@moe.edu.sg, matthew_ong@moe.gov.sg, lye_sze_yee@moe.gov.sg, lee_tat_leong@moe.edu.sg, xu_weiming@moe.edu.sg, goh_giam_hwee@moe.edu.sg, ong_chee_wah@moe.edu.sg, ng_soo_kok@moe.edu.sg, , lim_ee-peow@moe.edu.sg, lim_chew_ling@moe.edu.sg, yeo_wee_leng@moe.edu.sg, kenneth.lim@nie.edu.sg

Abstract: We are teachers who have benefited from the Open Source Physics (Brown, 2012; Christian, 2010; Esquembre, 2012) community's work and we would like to share some of the computer models and lesson packages that we have designed and implemented in five schools grade 11 to 12 classes.

In a ground-up teacher-leadership (MOE, 2010) approach, we came together to learn, advancing the professionalism (MOE, 2009) of physics educators and improve students' learning experiences through suitable blend (Jaakkola, 2012) of real equipment and computer models where appropriate . We will share computer models that we have remixed from existing library of computer models into suitable learning environments for inquiry of physics customized (Wee & Mak, 2009) for the Advanced Level Physics syllabus (SEAB, 2010, 2012).

We hope other teachers would find these computer models useful and remix them to suit their own context, design better learning activities and share them to benefit all humankind, becoming citizens for the world.

This is an eduLab (MOE, 2012b; Wee, 2010) project funded by the National Research Fund (NRF) Singapore and Ministry of Education (MOE) Singapore.

PACS: 01.50.H-  91.10.-v  96.20.Jz  04.80.-y  96.20.Jz

**Table 1: Summary of schools leading in the research and implementation of the lessons with computer model**

| Lead school | Customized computer model | Original model author and sub-author codes* | Figure | Number of teachers | Number of students | Scaling up in other schools |
|---|---|---|---|---|---|---|
| RVHS | Collision carts (ideal) | Francisco Esquembre Fu-Kwun Hwang* Andrew Duffy* | 1 | 3 | 242 | SRJC IJC |
| AJC | Collision carts (realistic) | Francisco Esquembre Fu-Kwun Hwang* Andrew Duffy* | 2 | 3 | 67 | On going |
| AJC | Falling magnet through solenoid | Francisco Esquembre | 3 | 8 | 198 | RVHS, SRJC |
| IJC | Ripple tank | Andrew Duffy Juan Aguirregabiria* Fu-Kwun Hwang* | 4 | 5 | 77 | YJC RVHS |
| YJC | Geostationary orbit | Francisco Esquembre Fu-Kwun Hwang* | 5 | 6 | 250 | On going |
| YJC | Field strength & potential | Andrew Duffy Fu-Kwun Hwang* | 6 | | | |
| YJC | Earth-Moon | Andrew Duffy Fu-Kwun Hwang* | 7 | | | |
| YJC | Kepler's 3rd law | Todd Timberlake | 8 | | | |
| SRJC | Superposition waves | Wolfgang Christian Fu-Kwun Hwang* | 9 | 7 | 145 | On going |

I. INTRODUCTION

We use a free authoring toolkit called Easy Java Simulation (EJS) (Esquembre, 2012) that allows ordinary teachers to create computer models as tools for interactive engagement (Adegoke, 2012; Hake, 1998) in physics education.

Building on open source codes shared by the Open Source Physics (OSP) community, and with help from the OSP community such as Fu-Kwun's NTNUJAVA Virtual Physics Laboratory (Hwang, 2010), we customized several Easy Java Simulation (EJS) computer models (Figure 1 to 9) that we hope many teachers will find useful. They are all downloadable and free to redistribute and use under creative commons attribution licenses from Digital Library in NTNUJAVA Virtual Physics Laboratory (Hwang, 2010) and our working Google site https://sites.google.com/site/lookang/.

In chronological order of implementation of the lessons, these are the lessons with computer models that we have used to interactively engage (Hake, 1998) our students, making physics come 'alive' and learn through meaningful play (Lee, 2012).

Aligned to our goal of scaling up (Dede, 2007) meaningful use of information and communications technology (ICT) into curriculum, assessment and pedagogy (MOE, 2008) we would briefly describe Figure 1 to 9 on these computer models. These computer models can be used and further customized (Wee & Mak, 2009) as scientific inquiry tools, suiting teachers' own "particular interests and educational points of view, and combine the use of a correct pedagogical approach with the sense of giving to it their own flavor" (Esquembre, 2002).

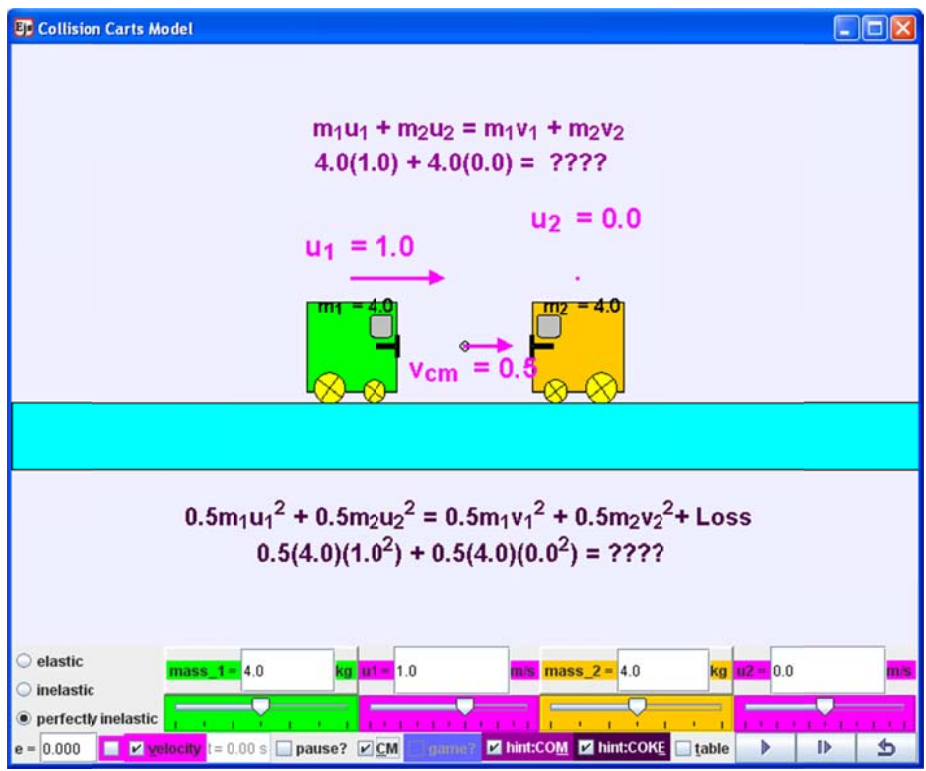

Figure 1. Collision carts (ideal) model (Wee, 2012b; Wee & Esquembre, 2008) derived from Francisco's original work (Esquembre, 2009) showing mathematical representations to illicit predictive thinking about the concepts.

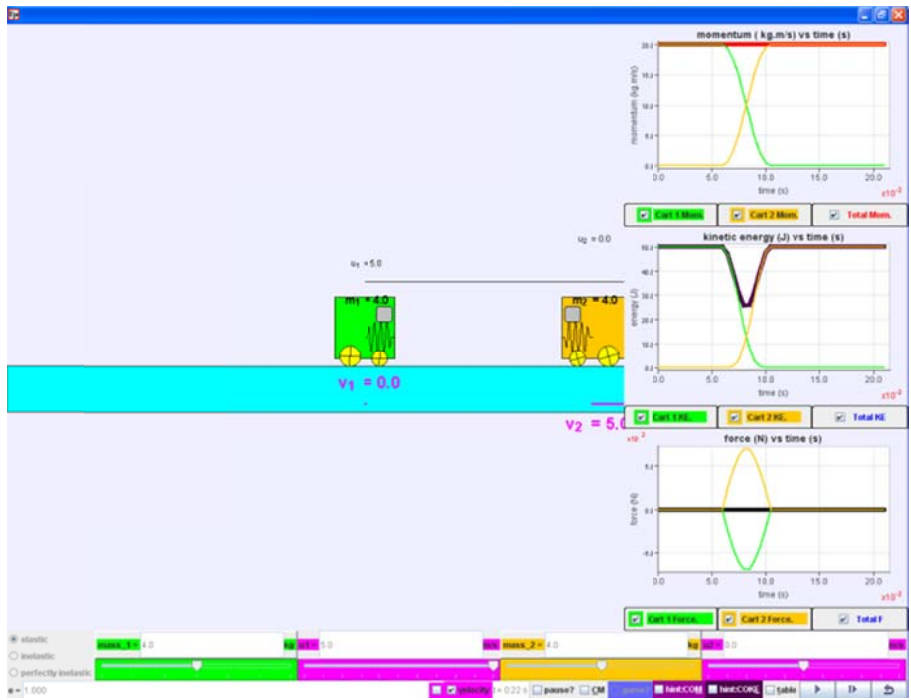

Figure 2. Collision carts (realistic) model (Wee, Esquembre, & Lye, 2012) derived from Francisco's original work (Esquembre, 2009) with 3 scientific graphs showing realistic spring modelled during collisions

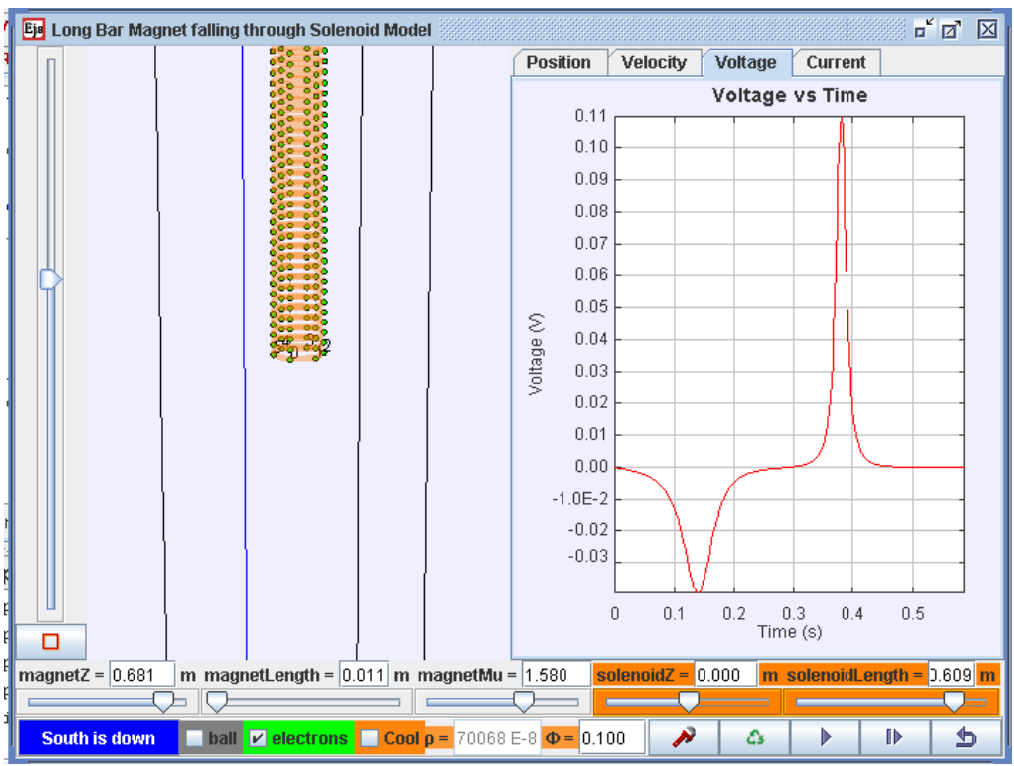

Figure 3. Falling magnet through solenoid model (Wee, Esquembre, & Lee, 2012) derived from Francisco's original work (Esquembre, 2010b) showing a long solenoid show and the resultant induced voltage as the short bar magnet falls through.

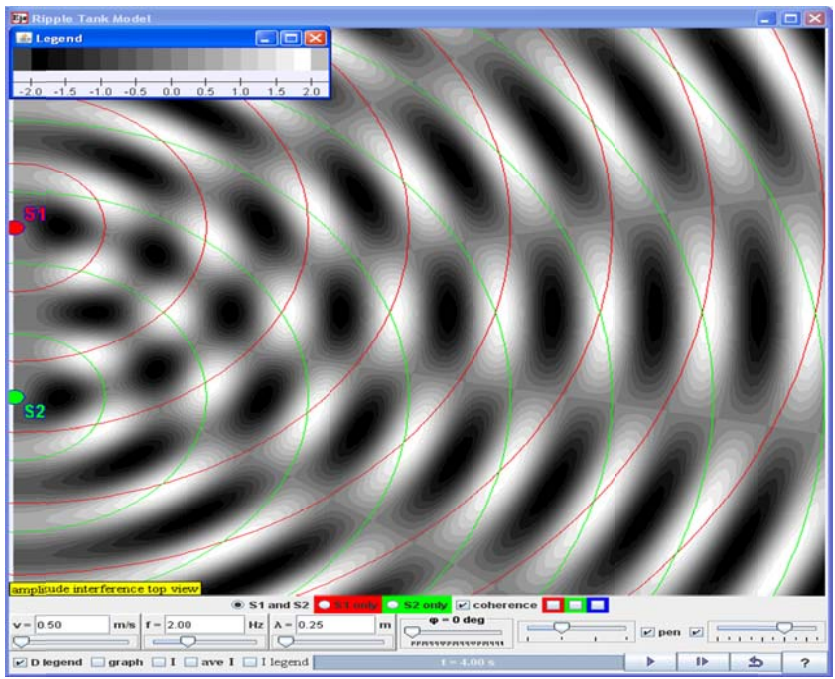

Figure 4. Ripple tank model (Wee, Duffy, Aguirregabiria, & Hwang, 2012) derived from Andrew's original work (Duffy, 2010) showing pen paper representation of crest and scalar field display of the interference pattern due to 2 point sources S1 and S2.

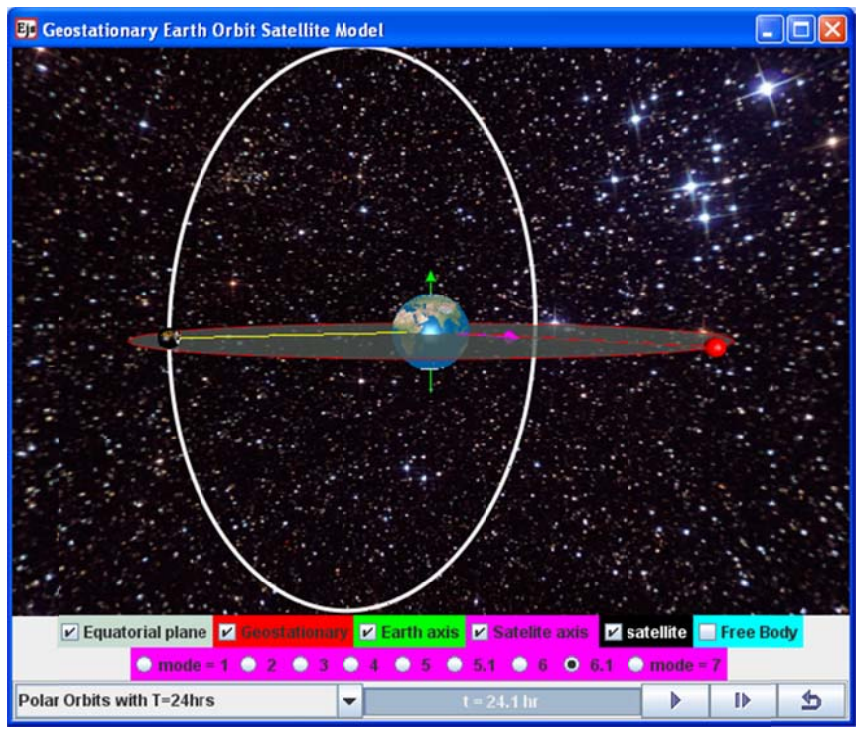

Figure 5. Geostationary orbit model (Wee, 2012a; Wee & Esquembre, 2010) derived from Francisco's original work (Esquembre, 2010a) showing a geostationary orbit (red) and a polar orbit (white)

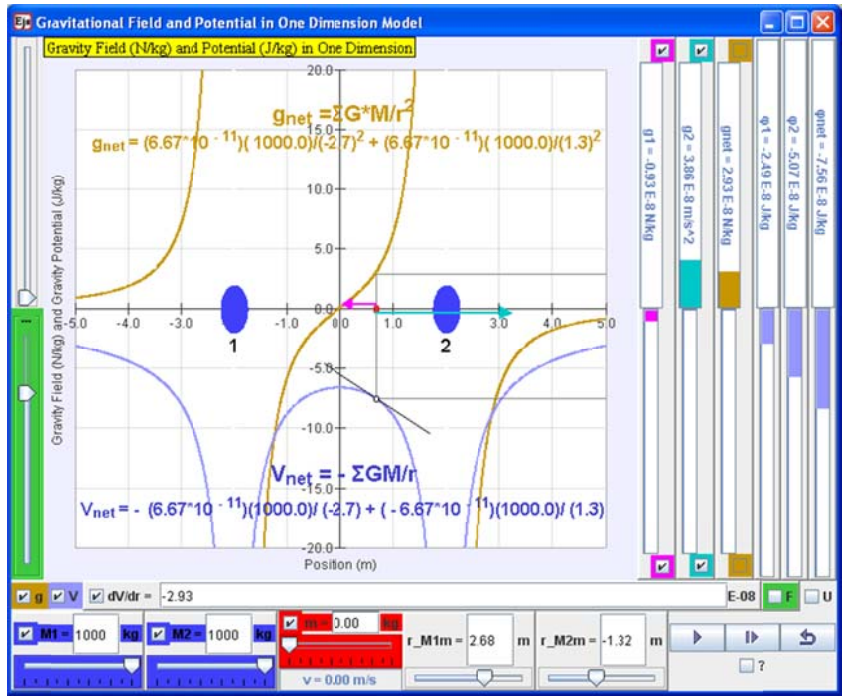

Figure 6.  Two mass model (Wee, Duffy, & Hwang, 2012a) derived from Andrew's original work (Duffy, 2009) showing a 2 mass system with gravitational and potential lines in 1 dimension

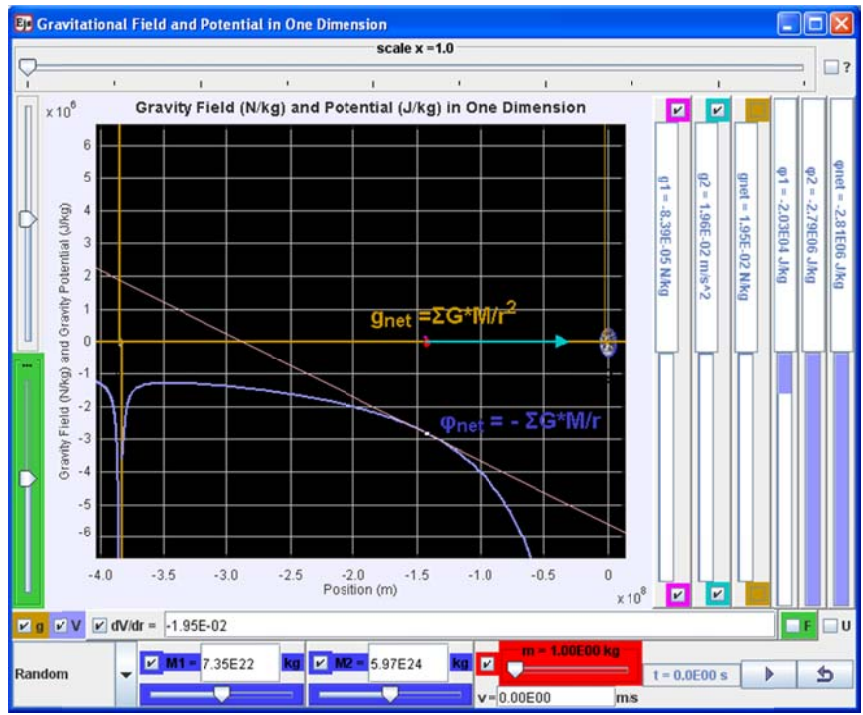

Figure 7.  Earth-Moon model (Wee, Duffy, & Hwang, 2012b) derived from Andrew's original work (Duffy, 2009) showing a 1 dimensional realistic model of the moon and earth system useful for exploring escape velocity concept.

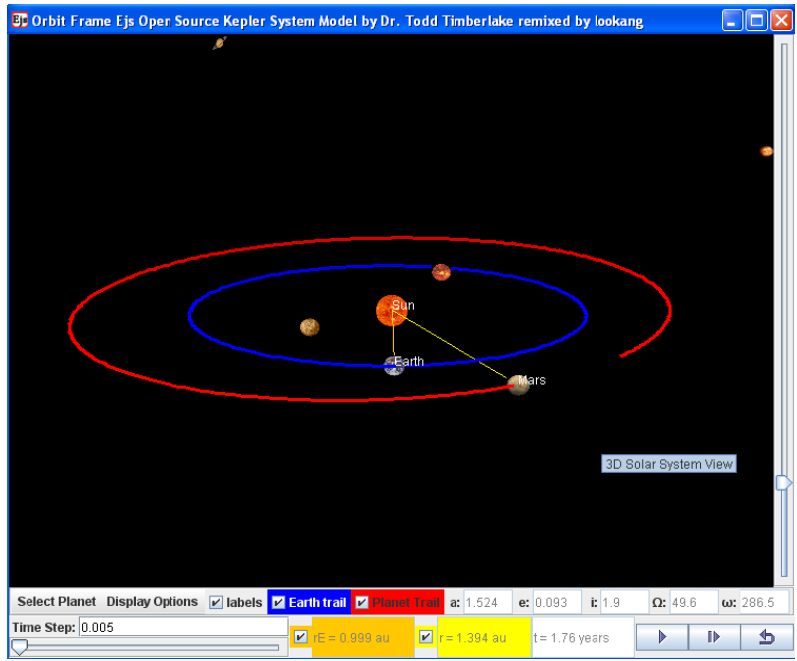

Figure 8.  Kepler's 3$^{rd}$ Law system model (Timberlake & Wee, 2011) derived from Todd's original work (Timberlake, 2010) showing earth and mars and their orbital trails for data collection of periods of planets.

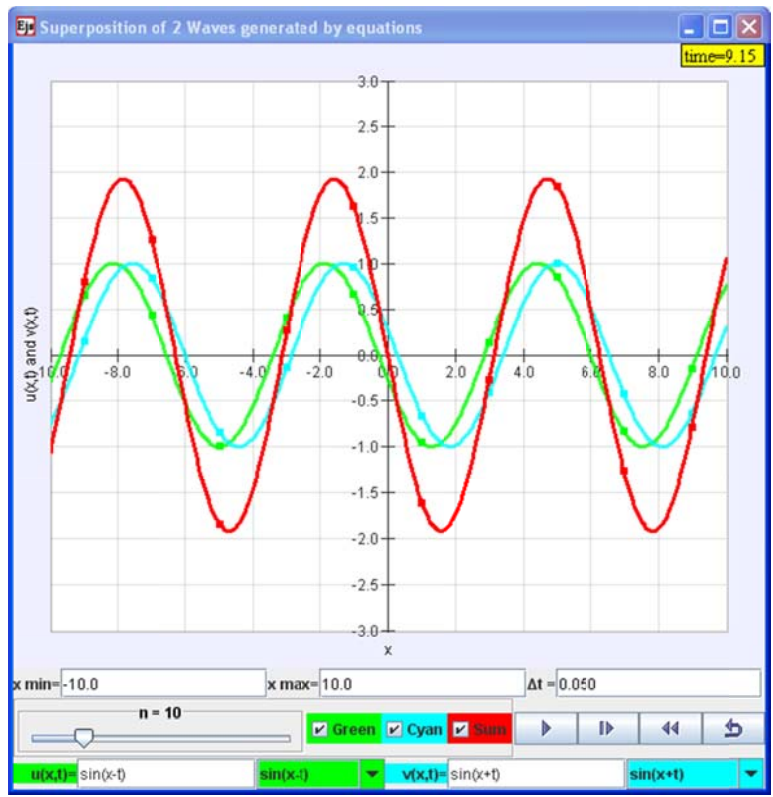

Figure 9.  Superposition of waves model (Wee, Christian, & Hwang, 2009) derived from Christian's original work (Christian, 2008) showing 2 functions and their resultant (red)

II. METHODS

**Table 2: Research methods used by school River Valley High School (RVHS) and Anderson Junior College (AJC)**

| School | Research Method | Students in experimental group | Students in control group | Teachers |
|---|---|---|---|---|
| RVHS | Lesson study | 242 | 0 | 3 |
| AJC | Experimental with pre-post test analysis | 67 | 62 | 3 |

A. *RVHS*

   1) *Setup*

   The students are seated in groups of 3 to 4 and are equipped with the worksheet and a laptop loaded with the computer model (Wee, 2012b; Wee & Esquembre, 2008). The questions in the worksheet were adapted from the Newtonian Tasks Inspired by Physics Education Research by Curtis J. Hieggelke, to suit the particular interest of the teachers, curriculum and the language of local students.

   2) *Worksheet*

   The worksheet (VI Appendix A) uses open ended scenarios of the 3 different collisions with lesson design influenced by predict-observe-explain (Liew & Treagust, 1995) strategy, to allow students to discuss and collaboratively decide on the most appropriate answers to consolidate and extend their understanding that can be simulated using the computer model.

   3) *Limitation of study*

   Using the lesson study approach, limitations of the study include reliance on teachers' subjective observation and difficulties in reviewing the large amount of video footage of the lesson.

B. *AJC*

   1) *Setup*

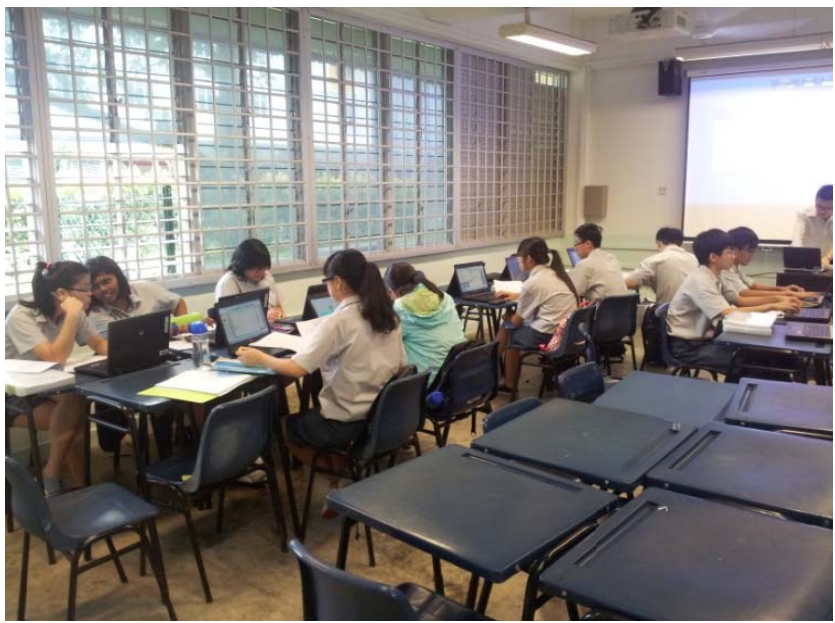

Figure 10. Lesson classroom seating arrangement in AJC. Each student uses a personal laptop with the Collision carts (realistic) computer model aided with inquiry worksheet questions modified from Physics by Inquiry (McDermott, Shaffer, & Rosenquist, 1995)

   The students are seated in groups of five and are equipped with the worksheet and a laptop loaded with the computer model (Wee, Esquembre, & Lye, 2012). The inquiry questions in the worksheet were adapted from the Physics by Inquiry (PBI) questions (McDermott, et al., 1995), to suit the content of local curriculum and language.

   2) *Worksheet*

   The worksheet (VII Appendix B) context of 2 gliders aims to promote conceptualization of Newton's First Law (uniform motion in frictionless surfaces) and Newton's Third Law of Motion (varying contact forces present during collision only, equal and opposite and on different bodies).

The worksheet focuses only on the example of collision of gliders and pre-post test questions are designed to lead students to conceptualize the same Newton's Three Laws in multiple representations (Wong, Sng, Ng, & Wee, 2011). Students are asked to interpret the *f-t* and *p-t* graph. They are also required to draw vector diagrams of forces acting on the gliders at different instants. The questions in the worksheet aim to improve student's mental construction of concepts and the process of knowledge construction is the focus of the lesson, guided by teacher facilitated discussion.

*3) Limitation of study*

Using the classical pre-test post-test research approach, limitations of the study include students not completing the pre/post-test to their abilities and inability to randomly divide into equivalent groups as classes are already pre-defined by school practices.

III. DATA AND FINDINGS

*A. RVHS*

Some other teachers went into the classroom to study the lesson and these are some of their observations.

We include excerpts of the lesson study notes by the teachers to give some themes and insights into the conditions and processes during the laboratory lessons.

*1) Need for well scaffolded inquiry activities*

"Students have difficulty because there weren't enough info given, no masses, no velocities. Not used to such open cases (tutorials – all open and shut cases). "

*2) Computer model can support inquiry activities*

"Students are not easily convinced that two moving objects colliding together come to a complete stop."

"(TH, YS and B) When faced with 2 contrasting theories, the more vocal or confident but wrong student (TH) was able to convince the other student (YS) of his answer. After the simulation, both appeared to quickly come to the right conclusion."

"They did not return to the previous wrong answer to look at why it was wrong. Once the answer is revealed, students tend to just focus on theories which fit the answers; regardless of their own feel that they feel something is wrong."

*B. AJC*

**Table 3: Experimental and control group comparison, where ↑ 0 ↓ represents improved, no change and deteriorated in post test scores respectively.**

|  | Experimental Group | | | Control Group | | |
|---|---|---|---|---|---|---|
| No. of students participated | 62 | | | 67 | | |
| No. of students used in <g> | 45 | | | 45 | | |
| No. of students who improved | 44 | | | 41 | | |
| Average Pre test score Max = 11 marks | $6.03 \pm 2.31$ | | | $6.74 \pm 2.10$ | | |
| Average Post test score Max = 11 marks | $7.60 \pm 2.26$ | | | $7.69 \pm 1.83$ | | |
|  | ↑ | 0 | ↓ | ↑ | 0 | ↓ |
| Newton's First law | 55% | 25% | 20% | 32% | 35% | 33% |
| Newton's Third law | 60% | 20% | 20% | 60% | 30% | 10% |
| % of students improved | 71% | | | 61% | | |

Table 3 suggests a higher percentage (55%) of students in the experimental group improved in the Newton's 1st Law than control group 32% while a fairly equivalent number of students improved in Newton's 3rd Law. Overall, the percentage of students that improved is (71- 61) = 10% higher in the experimental group compared to the control group. The increase in post-pre test scores is higher (1.56±2.93 vs 0.95±1.78) but the post scores are fairly equivalent (7.60 ± 2.26 and 7.69 ± 1.83). The standardized mean difference of the experimental over the control group is $\frac{1.56-0.95}{2.93} = 0.21$ which is medium in effect.

The average test scores in percentage (Figure 11) suggest higher score in Newton's 1st Law as well.

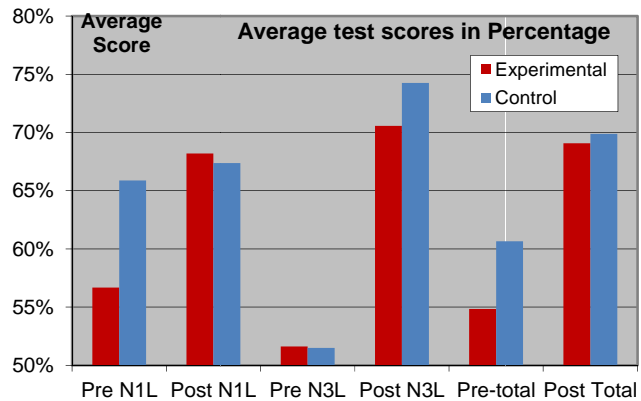

Figure 11. Average test scores in percentage of the 11 marks pre-post test with newton 1$^{st}$ law and 3$^{rd}$ law and total scores

Using pre-test scores plotted versus normalized gains (Figure 12), the trend of higher normalized gains $<g> = \frac{<\%post> - <\%pre>}{100 - <\%pre>}$ across 0 to 7 out of maximum 11 marks range of pre-test scores for the experimental group the emerged as well.

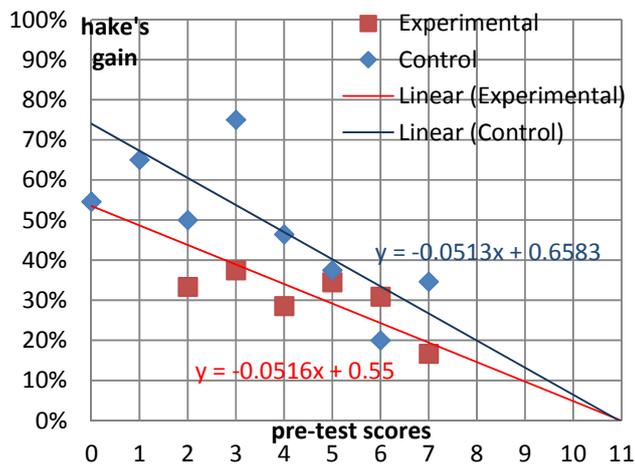

Figure 12. Graph of pre-test scores of both experimental (N=45) and control (N=45) groups versus hake's normalized gain $<g> = \frac{<\%post> - <\%pre>}{100 - <\%pre>}$ of students for the maximum scores of 11 marks with marks 8 to 11 omitted from analysis due to negative hake's gain but it does not adversely affect the overall trend.

IV. DISCUSSIONS

A. RVHS

We include excerpts from the qualitative survey results and informal interviews with the students to give some themes and insights into the conditions and processes during the laboratory lessons. Words in brackets < > are added to improve the readability of the qualitative interviews.

*1) Active and interactive engagment is key to learning*

"I felt good learning this way because it facilitated and encouraged discussion in groups, and ultimately allowing us to have deeper impression of certain concepts. It was a good experience and I feel I gained more from this lesson than from regular <less interactive> classes. I feel that students should be given the link to download the tool after lessons for interactive learning even at home, because this gives students a more interesting way to revise certain topics".

"The lesson was one of the best methods to actually be able to experiment and witness firsthand the results of different kinds of collision and is thus pretty good."

Oral Presentation [PS.02.09.a] The World Conference on Physics Education 1-6 July 2012

"I think we need more demonstration videos or java programs because people like me are visual learners and an animation will make learning clearer."

*2) Need good computer model design*

"The computer model design was brilliant and I enjoyed and understood the concept better with it."

"The applet provided an easy to understand interface and allowed for immediate understanding.

*3) Need stronger scaffolds at the beginning of lesson*

"The teachers should use the program to demonstrate and explain at the same time. Not letting us explore, and "wonder around".

"Explain the concepts first."

"It only feels good when our prediction and observations matched. Otherwise we were most of the time confused."

## B. AJC

*1) Hake's Gain and the rationale for removing data with high pre-test*

Several students in both groups obtained negative Hake's gain i.e. scored lower in the post test. We removed these students from the Hake's gain analysis because of several reasons.

  *a) increased Hake's gain senstivity at higher pre-test scores*

We speculate the increased sensitivity at higher pre-test scores say 8 marks, to score say lower say, 6 marks, which is computed as $= \frac{post-pre}{max-pre} = \frac{6-8}{11-8} = -200\%$, raised concerns about the validity of the test questions and students attitudes towards completing the tests.

  *b) Students making wild guesses during post test*

We speculate some students are making wild guess during the post tests suggested from interviews and absurdly short quiz time clocked online and perhaps also the lack of challenge from same questions in pre-post test.

Thus, marks 8 to 11 (Figure 12) were omitted from analysis resulting in our Hake's gain <g> analysis from 0 to 7 marks.

Although the Hake's gains <g> for the various pre test scores are plotted and a general trend emerged from the data that suggests Hake's gains of the students in the Experimental group are higher than the control group, from the linear fit (Figure 12) line.

Thus, with both medium effect standardized mean difference of 0.21 from experimental (N=62) and control (N=67) groups and the normalized gain analysis of the trend lines experimental (N=45) and control (N=45) groups and triangulated with interviews with students and teachers, the evidences suggest students did benefit from the experimental inquiry based lesson achieving deeper learning than their peers in traditional less interactive classrooms.

## V.  CONCLUSIONS

The 9 computer models derived from the Open Source Physics digital library are shared briefly giving credits to the original authors and sub-authors, so that ordinary teachers like us are able to stand on the shoulder of OSP giants and further customize our computer models to suit own syllabus and learning context.

## A. RVHS

General feedback from students and teachers has been relatively positive, thus the teachers will be scaling up (Dede, 2007) the use of other computer models with sound pedagogical approach.

## B. AJC

A medium effect of standardized mean difference of 0.21 from experimental (N=62) and control (N=67) and the higher normalized gain analysis from experimental (N=45) and control (N=45) across pre-test scores suggests that students who have benefitted from the inquiry based lesson can achieve deeper learning than their peers in traditional classrooms. In addition, general feedback from the students has been relatively positive, triangulated from students' survey and focus group interviews, reflections by teachers.

We hope more teachers will find the simulation useful in their own classes and further customized them so that others can act more intelligible (Juuti & Lavonen, 2006) with them, benefiting all humankind, becoming citizens for the world.

ACKNOWLEDGEMENT

We wish to acknowledge the passionate contributions of Francisco Esquembre, Fu-Kwun Hwang, Wolfgang Christian, Andrew Duffy, Todd Timberlake and Juan Aguirregabiria for their ideas and insights in the co-creation of interactive simulation and curriculum materials.

This research is made possible thanks to the eduLab project NRF2011-EDU001-EL001 Java Simulation Design for Teaching and Learning, (MOE, 2012b) awarded by the National Research Foundation (NRF), Singapore in collaboration with National Institute of Education (NIE), Singapore and the Ministry of Education (MOE), Singapore.

We also thank MOE for the recognition of our research on computer model lessons as a significant innovation in 2012 MOE Innergy (HQ) GOLD Awards (MOE, 2012a) by Educational Technology Division and Academy of Singapore Teachers.

Any opinions, findings, conclusions or recommendations expressed in this paper, are those of the authors and do not necessarily reflect the views of the MOE, NIE or NRF.

Oral Presentation [PS.02.09.a] The World Conference on Physics Education 1-6 July 2012MOE. (2010). An Introduction to PLCs Retrieved 01 December, 2010, from http://www.academyofsingaporeteachers.moe.gov.sg/cos/o.x?c=/ast/pagetree&func=view&rid=1069395

MOE. (2012a). MOE Innergy Awards: MOE Innergy (HQ) Awards Winners : Gold Award :Educational Technology Division and Academy of Singapore Teachers: Gravity-Physics by Inquiry Retrieved 25 May, 2012, from http://www.excelfest.com/award

MOE. (2012b). Press Releases: eduLab at the Academy of Singapore Teachers (eduLab@AST) to Bring Ideas into Practice Retrieved 25 May, 2012, from http://www.moe.gov.sg/media/press/2012/03/edulab-at-the-academy-of-singa.php

SEAB. (2010). Physics Higher 1 2011 8866. Retrieved from GCE A-Level Syllabuses Examined in 2011 website: http://www.seab.gov.sg/aLevel/20102011Syllabus/8866_2011.pdf

SEAB. (2012). 9646 Higher 2 PHYSICS (2013). Retrieved from GCE A-Level Syllabuses Examined in 2013 website: http://www.seab.gov.sg/aLevel/2013Syllabus/9646_2013.pdf

Timberlake, T. (2010). Kepler System Model 1.0. from http://www.compadre.org/Repository/document/ServeFile.cfm?ID=9757&DocID=1451

Timberlake, T., & Wee, L. K. (2011). Ejs Open Source Kepler 3rd Law System Model Java Applet 1.0. from http://www.phy.ntnu.edu.tw/ntnujava/index.php?topic=2225.0

Wee, L. K. (2010, 03 November). eduLab mass briefing on possible ideation options for eduLab projects sharing on Easy Java Simulation and Tracker. *Jurong Junior College*, 2010, from http://weelookang.blogspot.com/2010/10/edulab-mass-briefing-at-jurong-junior.html

Wee, L. K. (2012a). Geostationary Earth Orbit Satellite Model, from http://www.compadre.org/Repository/document/ServeFile.cfm?ID=11775&DocID=2634 & http://www.compadre.org/osp/document/ServeFile.cfm?ID=11775&DocID=2634&Attachment=1 (public download)

Wee, L. K. (2012b). One-dimensional collision carts computer model and its design ideas for productive experiential learning. *Physics Education, 47*(3), 301.

Wee, L. K., Christian, W., & Hwang, F.-K. (2009). Ejs Open Source Superposition of 2 Waves generated by equations, from http://www.phy.ntnu.edu.tw/ntnujava/index.php?topic=906.0

Wee, L. K., Duffy, A., Aguirregabiria, J., & Hwang, F.-K. (2012). Ejs Open Source Ripple Tank Interference Model java applet, from http://www.phy.ntnu.edu.tw/ntnujava/index.php?topic=2408.0

Wee, L. K., Duffy, A., & Hwang, F.-K. (2012a). Ejs Open Source Gravitational Field & Potential of 2 Mass Java Applet, from http://www.phy.ntnu.edu.tw/ntnujava/index.php?topic=1921.0

Wee, L. K., Duffy, A., & Hwang, F.-K. (2012b). Ejs Open Source Gravitational Field & Potential of Earth and Moon Java Applet, from http://www.phy.ntnu.edu.tw/ntnujava/index.php?topic=1924.0

Wee, L. K., & Esquembre, F. (2008). Ejs open source java applet 1D collision carts Elastic and Inelastic Collision, from http://www.phy.ntnu.edu.tw/ntnujava/index.php?topic=831.0

Wee, L. K., & Esquembre, F. (2010). Ejs Open Source Geostationary Satellite around Earth Java Applet requires Java 3D and Runtime. from https://sites.google.com/site/lookang/edulabgravityearthandsatelliteyjc/ejs_EarthAndSatelite.jar?attredirects=0&d=1 & http://www.phy.ntnu.edu.tw/ntnujava/index.php?topic=1877.0 (requires Registration to download)

Wee, L. K., Esquembre, F., & Lee, T. L. (2012). Ejs Open Source Long Magnet Falling Through solenoid Model Java Applet by LTL, from http://www.phy.ntnu.edu.tw/ntnujava/index.php?topic=2399.0

Wee, L. K., Esquembre, F., & Lye, S. Y. (2012). Ejs open source java applet 1D collision carts with realistic collision from http://www.phy.ntnu.edu.tw/ntnujava/index.php?topic=2408.0

Wee, L. K., & Mak, W. K. (2009, 02 June). *Leveraging on Easy Java Simulation tool and open source computer simulation library to create interactive digital media for mass customization of high school physics curriculum.* Paper presented at the 3rd Redesigning Pedagogy International Conference, Singapore.

Wong, D., Sng, P. P., Ng, E. H., & Wee, L. K. (2011). Learning with multiple representations: an example of a revision lesson in mechanics. *Physics Education, 46*(2), 178.
AUTHOR

Parallel Session 02.09|Date & Time: 02.07.2012 / 13:00 - 14:30|Hall: D403 (3rd Floor)    12/26

| | |
|---|---|
| 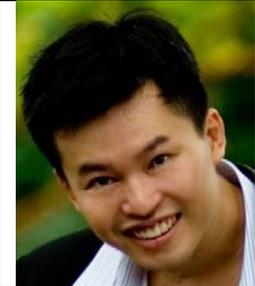 | Loo Kang Lawrence WEE is currently an educational technology specialist at the Ministry of Education, Singapore. He was a junior college physics lecturer and his research interest is in Open Source Physics tools like Easy Java Simulation for designing computer models and use of Tracker. |
| 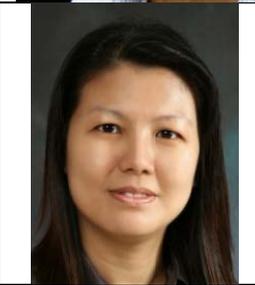 | Ai Phing LIM is currently a teacher in River Valley High School, Singapore. She has over 14 years of teaching grade 11 and 12 experience and has Masters in Science Education. |
| 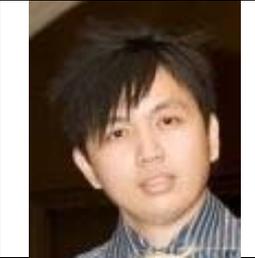 | Khoon Song Aloysius GOH is currently a physics teacher in Anderson Junior College. His academic and professional interests include the appropriate use of ICT to enhance learning and feasibility of organization management theories in Singapore's public school system. |
| 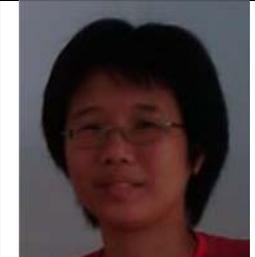 | Sze Yee is currently an educational technology officer in Ministry of Education, Singapore. She is a trained Physics Teacher and had taught both Physics and science in secondary and primary schools. She is now working on modifying the Open Source Physics Simulations for physics-related topics in primary school. |
| 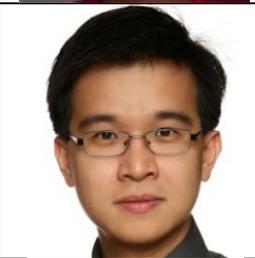 | Tat Leong LEE is currently the Head of Department for Education Technology in River Valley High School, Singapore. He is a high school Physics teacher, with 10 years of teaching experience. He has been using Open Source Physics (OSP) tools as early as 2006 (Tracker and Easy Java Simulations). |
| 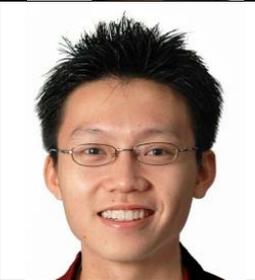 | Weiming XU is currently the Acting Subject Head for Educational Technology in River Valley High School, Singapore. He is a high school Physics teacher with a passion and interest in integrating multiple modes of representation of information in the teaching of Physics to provide authentic and meaningful learning experiences. |
| 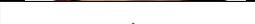 . | Giam Hwee Jimmy GOH is currently the Head of Science Department in Yishun Junior College, |
| 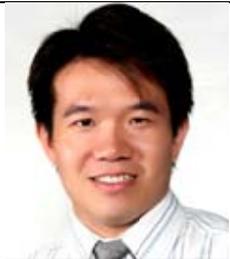 | Singapore. He teaches Physics to both year 1 and 2 students at the college and advocates inquiry-based science teaching and learning through effective and efficient means. |
| 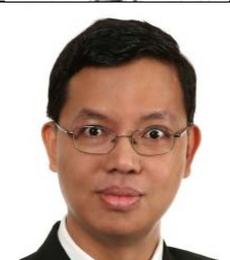 | Chee Wah ONG is currently a senior teacher teaching in Innova Junior College, Singapore. He has over 16 years of experience teaching grade 11 and 12 and has a Master degree in Science. |
| 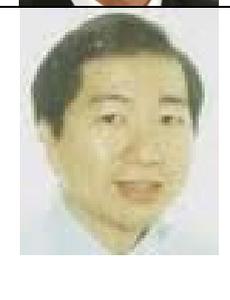 | Soo Kok, NG is currently teaching in Innova Junior College. He has 23 years of teaching experience and a keen advocate of experiential learning. |
| 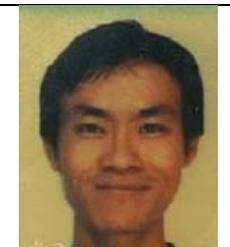 | Ee Peow LIM is currently teaching in Anderson Junior College, Singapore. He is leading a Physics ICT Resource Team of teachers. Before that, he obtained a distinction in pre-service teaching practicum with his creative teach methods. |
| 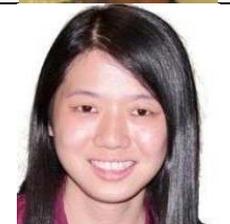 | Chew Ling LIM is currently teaching Physics in Serangoon Junior College, Singapore. She has 7 years of teaching experience. |
| 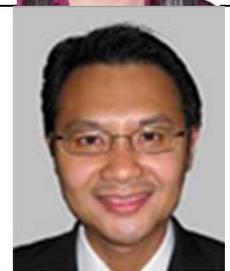 | Wee Leng Joshua YEO is currently teaching Physics in Serangoon Junior College, Singapore. He is also one of the College's ICT Mentor spearheading the initiative to create a critical mass of teacher advocates or champions to develop and cascade effective ICT practices. |

| 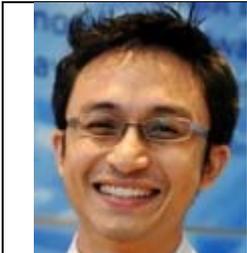 | Matthew ONG is currently an educational technology officer in Ministrhy of Education, Singapore. He has experience teaching in the grade 1 to 6. |
|---|---|
| 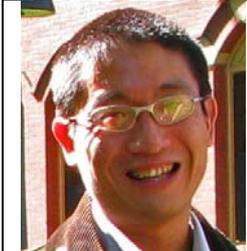 | Kenneth Y T LIM is a Research Scientist at the Office of Education Research, National Institute of Education. His present research interests are in the affordances for learning that immersive environments offer. Through his research, he is developing a theory of learning around the concept of Disciplinary Intuitions. |

VI. APPENDIX A: WORKSHEET WITH SUGGESTED ANSWERS BY RVHS

Subject: Physics  Time: 1h 15 min
Level: A-Level
Worksheet Title: P06 – Collisions between two bodies – Virtual Laboratory

Apparatus List
| 01 × laptop |
|---|

In this practical you will investigate the dynamics of collisions with TIPERs worksheets using the easy Java simulation ejs_users_sgeducation_lookang_Momentum1D2010web02.jar.

The following sections consist of various collision scenarios.

Read through the context carefully before making an educated guess as to the outcome. Explain your reasoning.

Finally, run the simulation to verify your prediction.

Are your predicted outcome and the simulated outcome identical? If they are not, explain the discrepancy.

How to use the Virtual Laboratory

Select the type of collision by clicking the radiobutton .
Key in the masses of the cart 1 and press the enter key. Repeat for cart 2
Key in the initial velocity of cart 1 and press the enter key. Repeat for cart 2.
Click the play button 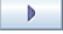 to start the simulation.
Reset the simulation by clicking on 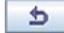 reset button.
You may wish to explore other features such as graphs in your own free time.
E.g.: *e* is the coefficient of restitution and it is the ratio of speeds after and before an impact, taken along the line of the impact (i.e. a measure of how much kinetic energy is lost).

**(a)** Carts A and B are shown just before they collide.

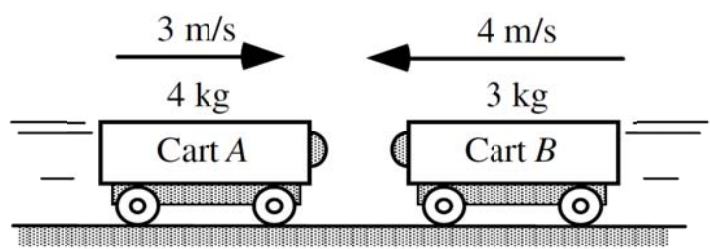

*No other information is given. Don't Ask.* ☺

Four students discussing this situation make the following contentions:

Eugene: "*After the collision, the carts will stick together and move off to the left. Cart B has more speed, and its speed is going to determine which cart dominates in the collision.*"

Sean: "*I think they'll stick together and move off to the right because Cart A is heavier. It's like when a heavy truck hits a car: The truck is going to win no matter which one's going fastest, just because it's heavier.*"

Thomas: "*I think the speed and the mass compensate, and the carts are going to be at rest after the collision.*"

Meili: "*The carts must have the same momentum after the collision as before the collision, and the only way this is going to happen is if they keep the same speeds. All the collision does is change their directions, so that Cart A will be moving to the left at 3 m/s and Cart B will be moving to the right at 4 m/s.*"

Which, if any, of these four students do you agree with?

Eugene_____ Sean _____ Thomas _____ Meili _____ None of them______

Explain.

Answer: None of these contentions is correct. We do not have enough information to determine the velocity of either cart after the collision. Momentum will be conserved for the collision, but this could happen in a number of ways, such as the carts sticking together and remaining at rest, or the carts bouncing off one another. What actually happens depends on the construction of the carts and on the material of the surfaces

…………………………………………………………………………………………….

**(b)** Two identical carts traveling in opposite directions are shown just before they collide. The carts carry different loads and are initially travelling at different speeds. The carts stick together after the collision.

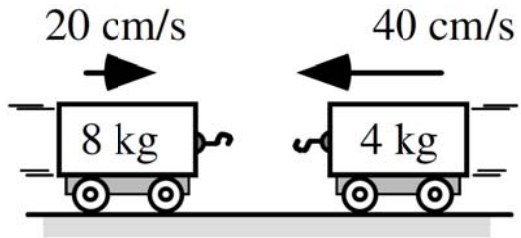

Three physics students discussing this situation make the following contentions:

Sherwin: "*These carts will both be at rest after the collision since the initial momentum of the system is zero, and the final momentum has to be zero also.*"

Sunny: "*If that were true it would mean that they would have zero kinetic energy after the collision and that would violate conservation of energy. Since the right-hand cart has more kinetic energy, the combined carts will be moving slowly to the left after the collision.*"

Steven: "*I think that after the collision the pair of carts will be traveling left at 20 cm/s. That way conservation of momentum and conservation of energy are both satisfied.*"

Which, if any, of these three students do you think is correct?

Sherwin _____   Sunny _____   Steven _____   None of them______

Please explain your reasoning.

Answer: Sherwin is correct. The momentum of the two carts are equal and opposite before the collision, so the total initial momentum is zero and the total final momentum has to be zero also.

**(c)**   In Case A, a *metal* bullet penetrates a wooden block.
In Case B, a *rubber* bullet with the same initial speed and mass bounces off an identical wooden block.

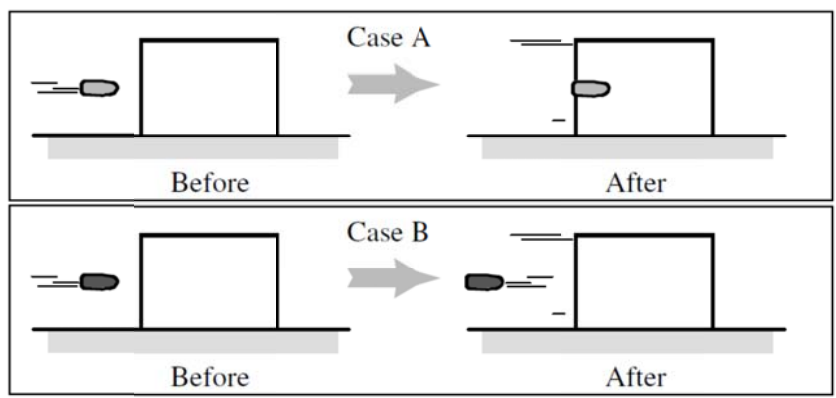

*No other information is given. Don't Ask.* ☺

Will the speed of the wooden block after the collision be greater in Case A, greater in Case B, or the same in both cases?

Explain.

Answer: Greater for B. The initial momentum in both cases is the same and points to the right. The final momentum of the bullet points to the right in Case A and to the left in Case B. Since the final momentum of the system consisting of the bullet and the block is the same as the initial momentum, and this final momentum is the vector sum of the momentum of the bullet and the momentum of the block, the momentum of the block must be greater in Case B.

Will the speed of the bullet in Case B after the collision be greater than, less than, or the same as the speed of the bullet just before the collision?

Explain.

Answer: Less than. The energy of the system containing both block and bullet cannot be greater after the collision than before. The initial energy is the kinetic energy of the bullet, and the final energy is the sum of the kinetic energies of the bullet and the block. Since the block has a non-zero final kinetic energy, the final kinetic energy of the bullet must be less than the initial kinetic energy of the bullet.

Half Way Check Point

For each of the earlier situations (a) to (c), answer the following questions.

List all the external forces exerted on the system.
Does the system have an initial momentum? Describe any changes in its total momentum.
Does the system experience a net impulse during the specified time period? Explain.

**(a)**

1. Assume friction is negligible.

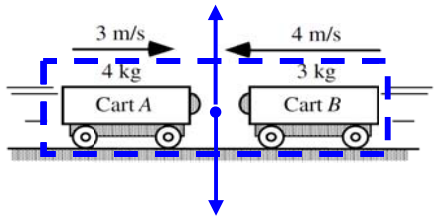

2. Initially, the system has zero momentum. The total momentum does not change with time. Or rather, the change in momentum is zero.

3. There is no net impulse delivered to the system. The gravitational and normal forces balance.

**(b)**

1. Assume friction is negligible.

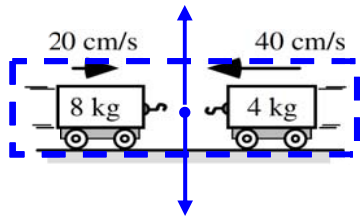

2. Initially, the system has zero momentum. The total momentum does not change with time. Or rather, the change in momentum is zero.

3. There is no net impulse delivered to the system. The gravitational and normal forces balance.

ooden

tion
oden

illet

**(c)**  1.  Assume friction is negligible.

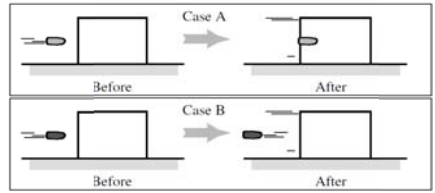

2. Initially, the system has momentum. The total momentum does not change with time. Or rather, the change in momentum is zero.

3. There is no net impulse delivered to the system. The gravitational force on the bullet causes a small vertical downward change in momentum of the bullet, which is negligible.

**Extending Your Understanding**

(e)    Two identical steel balls, *P* and *Q*, are shown at the instant that they collide.

The paths and velocities of the two balls before and after the collision are indicated by the dashed lines and arrows.

The speeds of the balls are same before and after collision.

For the questions below, use the directions indicated by the arrows in the direction rosette, or use *J* for no direction, *K* for into the page, *L* for out of the page, or *M* if none of these are correct.

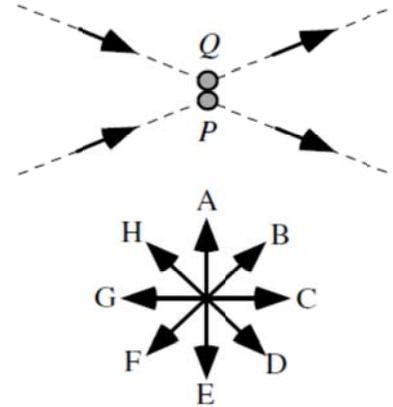

Which letter best represents the direction of the change in momentum for ball *Q*?

Explain.

Answer: ***A***. The change in velocity of ball Q is its final velocity minus its initial velocity, and is found by subtracting vectors as shown.

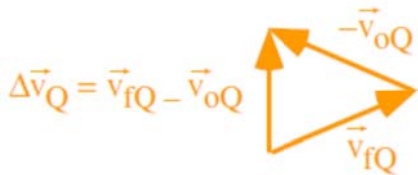

Which letter best represents the direction of the change in momentum for ball *P*?

Explain.

Answer: ***E***. The change in velocity of ball P is its final velocity minus its initial velocity, and is found by subtracting vectors as shown.

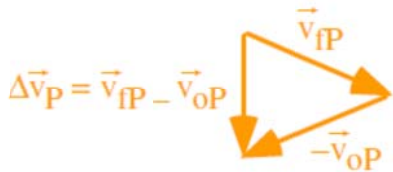

Choose the letter that best represents the direction of the initial momentum for the system of both balls *P* and *Q* before collision.

Explain.

Answer: **C**. The initial momentum of the system is the vector sum of the initial momentum of the individual balls. When added together, these momentum point to the right as shown.

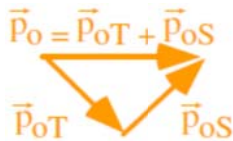

Choose the letter that best represents the direction of the final momentum for the system of both balls *P* and *Q* after collision.

Explain.

Answer: **C**. The final momentum of the system is the vector sum of the final momentum of the individual balls. When added together, these momentum point to the right as shown. Note that since momentum is conserved for this system, the final momentum is equal to the initial momentum.

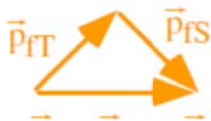

Choose the letter that best represents the direction of the impulse during this interaction for the system of both balls *P* and *Q*.

Explain.

Answer: **J**. There is no direction since there is no impulse on the system during the interaction. There are no external forces, and so no impulse and no change in momentum for the system.

VII. APPENDIX B: WORKSHEET WITH SUGGESTED ANSWERS BY AJC

Experiment 1

Glider A is launched towards and collides *inelastically* with a stationary glider B on a smooth plane. After the collision, glider A reverses direction. The mass of glider A, m is one fifth the mass of glider B.

Given the conditions above, attempt the Java simulation with different values of initial velocities.

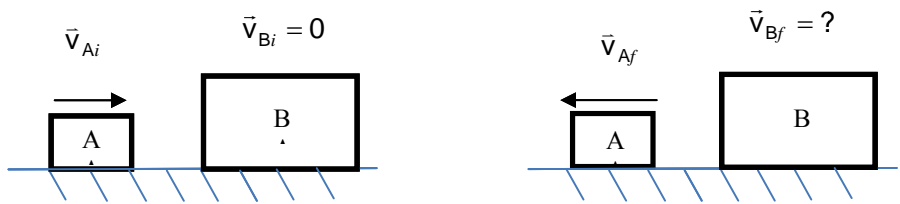

In the space provided, draw separate free-body diagrams for each glider and for the system S of the two gliders at an instant during the collision.

| er A | r B | |
|---|---|---|
| | | |

(B) How does the net force on glider A, FA, compare to the net force on glider B, FB, at this instant? Discuss both the magnitude and direction of the net force.

The magnitude of FA is ~~greater than~~ / equal to / ~~smaller than~~ the magnitude of FB.
The direction of FA is ~~same as~~ / opposite to the direction of FB.

(C) The F-t graph (Figure 1) below shows the net force FA acting on glider A during the collision. On Figure 1, sketch the variation with time t of net force FB acting on glider B.

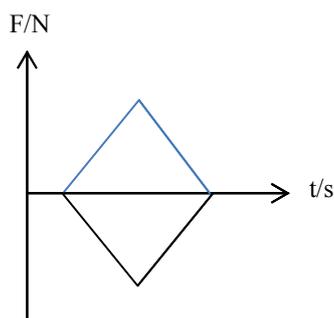

Figure 1

(A) What can you say about the net force acting on glider B
    (i)   before collision    zero
    (ii)  during collision   $F_B$ (equal to $F_A$)
    (iii) after collision    zero

Now, consider the time interval while the gliders are still in contact during the collision to be Δt. How does the product $F_A \Delta t$ compare to the product $F_B \Delta t$? Discuss this in terms of the magnitude and direction.

Since $F_A$ equal to $F_B$ in magnitude and opposite in direction, $F_A \Delta t$ and $F_B \Delta t$ are equal in magnitude and opposite in direction.

Apply Newton's second law (for constant mass) $F_{net} = m\frac{\Delta v}{\Delta t}$ to each of the gliders to compare the change in momentum (Δp=mΔv) of gliders A and B during the collision. Discuss both the magnitude and direction of the change in momentum.

The magnitude of the change in momentum of glider A is ~~greater than~~ / equal to / ~~smaller than~~ the magnitude of the change in momentum of glider B.
The direction of the change in momentum of glider A is ~~same as~~ / opposite to the direction of the change in momentum of glider B.

The area under an F-t graph represents the change in momentum of a body. Hence, on Figure 2, sketch corresponding momentum-time graphs for
glider A
glider B

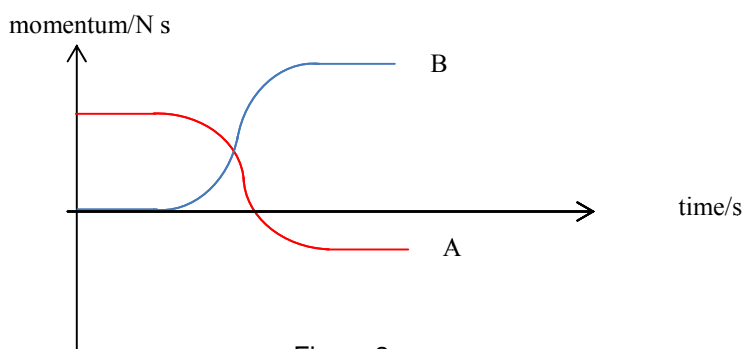

Figure 2

Describe the momentum-time graph for glider B after the collision.

The momentum-time graph for glider B after collision is a straight horizontal line.

Using your answers to D (iii) and H, what can you say about the velocity of glider B during this period?

The velocity of glider B during this period is constant. (No net force, no change in momentum)

Experiment 2

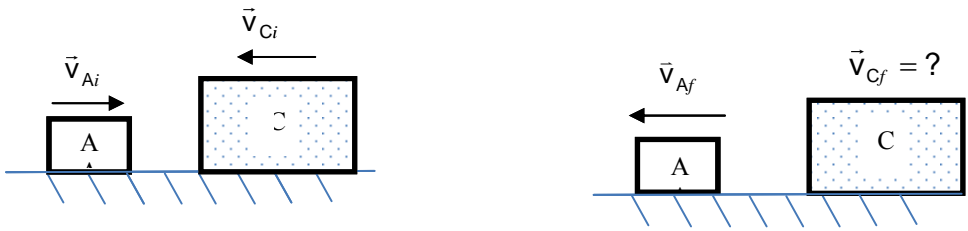

Glider A is now launched with a momentum of 200 kg m s-1 towards glider C which is moving in the opposite direction with a momentum of 50 kg m s¬-1. After the inelastic collision, glider A reverses direction. The mass of glider A is 25 kg and the mass of glider C is 40 kg. The coefficient of restitution e for this collision is 0.897.

(A)    In the space provided, draw separate free-body diagrams for each glider and for the system S of the two gliders at an instant during the collision.

| glider A | ler C | S |
|---|---|---|
|  | ⟶ $F_C$ |  |

How does the net force on glider A, $F_A$, compare to the net force on glider C, $F_C$, at this instant? Discuss both the magnitude and direction of the net force.

The magnitude of $F_A$ is ~~greater than~~ / equal to / ~~smaller than~~ the magnitude of $F_C$.
The direction of $F_A$ is ~~same as~~ / opposite to the direction of $F_C$.

Discuss the magnitude and direction of the change in momentum.

The magnitude of change in momentum of glider A is ~~greater than~~ / equal to / ~~smaller than~~ the magnitude of change in momentum of glider C.
The direction of change in momentum of glider A is ~~same as~~ / opposite to the direction of change in momentum of glider C.

Using values obtained from the Java simulation, fill in the final momentum of glider A and complete the momentum-time graph of glider C with an appropriate value.

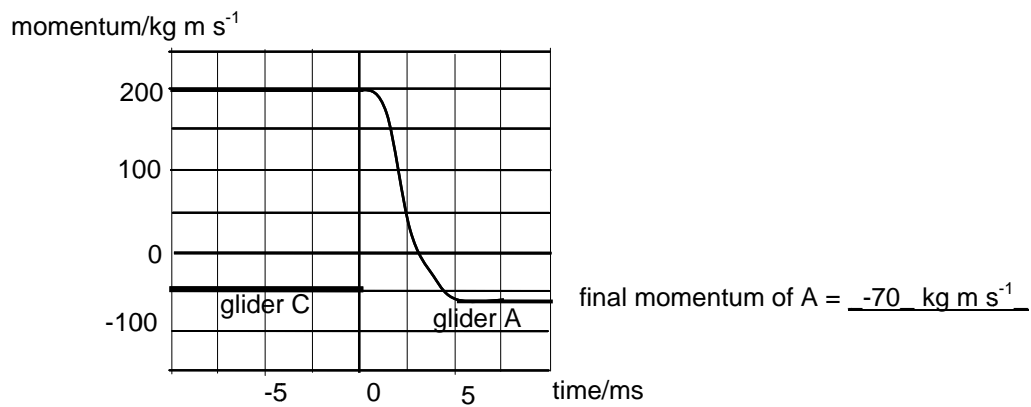

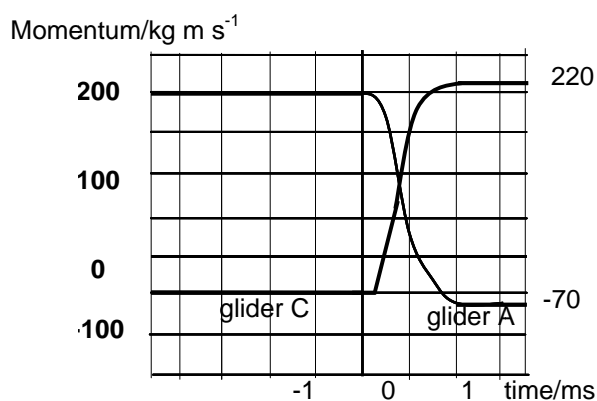

**Thinking Questions**
Comment on the velocities of the bodies after they collide elastically for the following situations:
1. 2 identical masses colliding.
2. A tennis ball incident on a wall
3. A bowling ball incident on a stationary table tennis ball.
Use the Java simulation to confirm your results.

Adapted from Tutorials in Introductory Physics    ©Prentice Hall, Inc.
Mc Dermott, Shaffer, & P.E.G., U. Wash.                                           First Edition, 2002